**Authors' Information**

*Hector Garcia* - Contrated Professor. Technical University of Madrid. E.U. Informática. Ctra. de Valencia Km. 7. E28031 Madrid. e-mail: hgarcia@eui.upm.es

*Carlos del Cuvillo* - Associate Professor. Technical University of Madrid. E.U. Informática. Ctra. de Valencia Km. 7. E28031 Madrid. e-mail: ccuvillo@eui.upm.es

*Diego Perez* -. Consultant. Technical University of Madrid. E.U. Informática. Ctra. de Valencia Km. 7. E28031 Madrid. e-mail: dperez@tdi.eui.upm.es

*Borja Lazaro* - Technician, Group leader. Technical University of Madrid. E.U. Informática. Ctra. de Valencia Km. 7. E28031 Madrid. e-mail: blazaro@eui.upm.es

# MODELING AND ANNOTATING THE EXPRESSIVE SEMANTICS OF DANCE VIDEOS

## Balakrishnan Ramadoss, Kannan Rajkumar

*Abstract: Dance videos are interesting and semantics-intensive. At the same time, they are the complex type of videos compared to all other types such as sports, news and movie videos. In fact, dance video is the one which is less explored by the researchers across the globe. Dance videos exhibit rich semantics such as macro features and micro features and can be classified into several types. Hence, the conceptual modeling of the expressive semantics of the dance videos is very crucial and complex. This paper presents a generic Dance Video Semantics Model (DVSM) in order to represent the semantics of the dance videos at different granularity levels, identified by the components of the accompanying song. This model incorporates both syntactic and semantic features of the videos and introduces a new entity type called, Agent, to specify the micro features of the dance videos. The instantiations of the model are expressed as graphs. The model is implemented as a tool using J2SE and JMF to annotate the macro and micro features of the dance videos. Finally examples and evaluation results are provided to depict the effectiveness of the proposed dance video model.*

*ACM Classification: J.5 [Arts and Humanities]: Performing Arts; H.5.4 [Information Interface and Presentation]: Hypertext/Hypermedia.*

## 1. Introduction

Dance data is essentially multimedia by nature consisting of visual, audio and textual materials. Dance video modeling and mining depends significantly on our ability to recognize the relevant information in each of these data streams. One of the most challenging problems here is the modeling of the dance video semantics such that the relevant semantics are consistent with the perception of the real world.

The classical and folk dances are the real cultural wealth of a nation. In India, the most important classical dances are *Bharathanatyam, Kadak, Kadakali, Kuchipudi* and *Manipuri* (Saraswathi, 1994). Traditionally, dance learners perform dance steps by observing the natural language verbal descriptions and by emulating the steps of the choreographers. Therefore, the properly annotated dance videos will help the present and future generations to learn dance themselves and minimize the physical presence of the choreographers.

Notations are used everywhere and are most important for the dancers to communicate the ideas to the learners. They use graphical symbols such as vertical lines, horizontal lines, dots, triangle, rectangle etc, to denote body parts' actions on paper. Labanotation (Hutchinson, 1954) and Banesh (Ann, 1984) have been the frontier notational systems to record the dance movements or dance steps. Many western dances are using Labanotation to describe dance steps. However, many choreographers still follow the traditional way of training their students using natural language descriptions, because of the very few recording experts and inherent complexity of reading and understanding the symbols. Moreover, all Indian dances have unique structure and no common notational structure exists, apart from wire-frame stick diagram representing a dance step. Due to lack of notations, it is evident that the complexity of modeling the dance video semantics is relatively high.

Since the dance steps were archived in paper form and many classical dances lack notations, this kind of archival of dance becomes impossible even today. With the advances in digital technologies (Dorai, 2002) nowadays, magnetic tapes and disks record dance presentations efficiently. But, searching a dance sequence from these collections is not efficient, because of the huge volume of video data. The solution is to build a dance video information system so as to preserve and query the different dance semantics like, dance steps, beyond the spatio-temporal characteristics of the dancers and their dance steps.

The dance video database system requires an efficient video data model to abstract the semantics of the dance videos. To be more precise, the dance video data model should:

- abstract the different dance video semantics such as dancers, dance steps, agents (i.e., body parts of the dancers), posture, speed of dance steps, mood, music, beat, instrument used, background sceneries and the costume. More importantly, the spatio-temporal characteristics of the dancers must be incorporated in the model;
- capture the structure of the dance videos such as shot, scene and compound scene abstracting the different components of the accompanying song.

This paper addresses two related issues: modeling the semantics of the dance videos and annotating the dance steps from the real dance videos. The dance video semantics model represents the different types of dance semantics in a simple, efficient and flexible way. The annotation tool manually annotates the semantics (as macro and micro features) for further query processing and video mining. The main contributions of this paper are as follows:

- We propose a generic video data model to describe the dance steps as video events;
- We introduce the Actor entity in order to store the event specific roles of a video object. That is, an actor entity describes the context dependent role of the video object;
- We introduce the Agent entity to describe the context dependent action that is associated with the actor entity;
- We develop a tool that implements the dance video model in order to annotate the different dance semantics.

The rest of the paper is organized as follows: Section 2 presents some related works on video data models. Section 3 describes the different semantics of the dance videos. The DVSM for the dance video is introduced in Section 4. Section 5 illustrates the implementation of the DVSM using Java technologies. The proposed video model is evaluated against a set of conceptual and semantic quality factors in Section 6. Finally, Section 7 concludes the paper.

## 2. Related Work

Video data modeling is an important component of the dance video database system, as it abstracts the underlying semantics of the dance. This section briefly reviews some of the existing video modeling proposals and discusses the applicability to dance videos.

Colombo(1999) classifies the content-based search as semantic level search (e.g. objects, events and relationships) and low level search (e.g. color, texture and motion). They call the corresponding systems as first and second generation visual information systems. Several key word based techniques are applied to semantic search models, such as OVID (Oomoto, 1993), AVIS (Adali, 1996), Layered model (Koh, 1999) and Schema less semantic model (Al Safadi, 2000). Second generation systems provide automatic tools to extract low level features and subsequently semantic search is performed. Some of these systems include, but not limited to QBIC, Virage, VisualSEEK, VideoQ, VIOLONE, MARS, PhotoBook, ViBE, and PictHunter (Smeulders, 2000;

Antani, 2002). However, these systems are either based on textual annotations or purely low level features, but not incorporating the other one.

In (Shu, 2000), Augmented Transition Network based semantic data model is proposed. The ATN models the video based on scenes, shots and key frames using strings as a sequence of characters. The string representation is used to model the spatial and temporal relationships of each object (moving and static) in a shot of the traffic video. Since the semantic features of dance videos are complex, the entire scene or shot cannot be abstracted in a single string.

Translucent markers, reflector costumes, special sensors and specialized cameras are used to capture and track human body parts' movements in some applications such as aerobics, traffic surveillance, sign language, news and sports videos (Vendrig, 2002). In order to record and analyze the dance steps of a dancer, based on this technique requires a special translucent markers or reflector costumes for the dancers. However, dancers do not prefer to use these costumes as these costumes hide the dancer's make-ups and costumes. Moreover, these markers and reflectors prohibit the realism, affect dancer's comfort as well as reduce the focus or concentration of the dancers. Hence, automatic analysis of dance steps to extract the semantics of the dance steps is very complex.

Recently, the extended DISIMA(Lei Chen, 2003) model expresses events and concepts based on spatio-temporal relationships among salient objects. However, the required dance video database model has to consider not only salient objects, but all objects such as instruments, costumes, background and so on.

Event based syntactic-semantic video model (we call it as, ESSVM) (Ahmet, 2004) proposes Actor entity to specify the context dependent role of a player in soccer sports. This model represents the events such as free kick, goal, penalty etc, in which player assumes different roles such as scorer, assist-maker etc. But in dance videos, the contextual information of the dance events has to be described at multiple levels like actor and agent, rather than at a single granularity of actor entity.

COSMOS7 (Athanasios, 2005) models objects along with a set of events in which they participate, events along with a set of objects and temporal relationships between the objects. This model does not model the temporal relationships between events and the contextual roles. It models the events at a higher level only like speak, play, listen etc, whereas dance video model needs more detailed level of event representation such as agents, their action, speed of action, associated song and so on.

## 3. The Semantics of Dance Videos

Generally, dance information is dominated by visual content such as steps, posture and costume and the accompanying audio such as song and music. Hence, dance videos are rich in semantics and provide ample scope for the efficient semantic retrieval and dance video mining. This section illustrates the song that accompanies the dance performance, the different dance video types and the features of the dance videos in detail.

### 3.1. Song Granularity

Dance video contains several dance steps representing each song. In the case of classical dance, it is simply a collection of songs choreographed on the stage or theatre with a single start-stop (Cheng, 2003) camera operation. On the other hand, in a movie dance, a movie contains several songs and for each song dance steps are choreographed by the dancers. A song in a movie may be recorded with multiple start-stop camera operations. For instance, an Indian movie will normally contain about five to six songs. Here, each dance step may represent a step from any of the Indian dances or a new step innovated by the choreographer. Further, it includes the presentation aesthetics such as mood, feelings, emotion and so on.

Song is composed of four parts: Introduction, Additional Introduction, Chores and Stanzas (Web of Indian Classical Dances, 2003). Depending on the type of a song, Additional Introduction and Chores may be optional. Each part has few lines of lyrics for which dance steps are choreographed. In the dance video hierarchy, a shot represents a dance step, a scene represents dance steps of any of the song parts which are recorded in the same location and a video clip represents dance steps of a song. Our DVSM will represent the semantics of one dance step as a dance event. Dance step is the unit of analysis in this paper.

### 3.2. Features of Dance Videos

There are two types of dance video features- macro features and micro features and are annotated by the human annotators at macro and micro levels (Forouszan, 2004) accordingly. Macro features are general properties of the dance that are event independent and micro features are the properties of the dance step. That is, micro features are spatio-temporal characteristics of the dancers while rendering the dance steps. Micro features can also be called as event dependent features.

Macro features(or Bibliographic features): Date of recording, time of recording, geographic origin of the dance, geographic origin of the dancers, sex, age, number of dancers in a dance, type of performance venue (such as theatre, open-air, etc), type of the accompanying song, type of accompaniment, type of musical instrument used and types of dance videos. The different dance videos are movie dance video, theatre dance video, folk dance video, classical dance video, street dance video and festival dance video. These macro features are independent of the dance steps and are common to all dances.

Micro features (Dance step specific features): Spatio-temporal features classify dance movement behavior which include: movement of one dancer in relation to another dancer, movement of a specific body part (such as eye, leg etc. Refer Appendix-A for a complete list) of a dancer in relation to another part of the body, movement path of the dance (such as circular, linear, serpentine and zigzag), distance between body parts of a dancer while performing a dance step and distance between dancers.

Hence, the proposed video model has to characterize a set of macro features and micro features that exist in the dance videos.

## 4. The Dance Video Semantics Model

Conceptual model abstracts the dance video data into a structure for later querying its contents and mining some interesting patterns. For efficient conceptual modeling, one should know how choreographers demonstrate a dance to the learners. They are the experts in describing the rhythmic steps to the audience. This section presents a generic dance video model that efficiently describes the dance steps. Every dance step is called as an event and the model represents dance events by a set of micro features. The model is generic in the sense that it is applicable to any type of dance videos. DVSM is an extension of ER (Chen, 1976) with object oriented features. The goal of the model is to describe a dance step as an event.

The main entities of the model are events, objects that participate in these events, actor entities that describe contextual roles of objects in the events, agent entities that represent the action of the actor and concept entities that model the cognitive and affective features of the dancers.

For example, consider a dancer object with name, age, address and all other event independent attributes. The same dancer assumes different roles throughout the dance video. That is, he becomes hero in one dance step, lover in another dance step and so on. Roles are defined as attributes of Actors. Some other examples of actors are heroine, leader, follower, group dancer, friend etc. These context specific object roles form separate actor entities, which all refer to the same dancer object. Although one would say that actor performs the action in an event, finer granularity is necessary as far as dance videos are concerned. Therefore, contextual data of the dancers have to be described in two levels. A particular dance step is characterized by the actions of the agents who belong to the actors. Spatio-temporal characteristics are part of the actors as well as agents. Hence, they are described as attributes of actors and agents.

Apart from the dancer object, DVSM may also represent the ordinary objects with a standard UML class diagram. Some of them are: speed of the action of an agent, instrument used and the posture of the actor. The graph meta-schema of the DVSM is depicted in Figure1.

The graphical notations used in DVSM are described as follows. A rectangle node refers to an entity or an object. A round rectangle node refers to a concept. A dotted rectangle node denotes an actor entity. A thick rectangle node shows an agent object. Event entity is modeled with a trapezoid. Attributes of entities and relationships are represented with oval nodes. Relationships are denoted with directed lines on which the name of the relationship is denoted. Relationships without their names represent the containment type.

The model is instantiated as a directed acyclic graph. The reason for choosing graphs is that it elegantly models repetition of dance steps and has matured as a graph database. If a dance step repeats after some time, it just requires another edge to point to the same node. A graph is formally defined as follows: Let $G = (V, E)$ be a directed acyclic graph, where V denotes set of vertices and E denotes set of directed edges. The different entity classes, events, actors, agents, concepts, and other basic classes become vertices of the graph. Similarly, the set of interaction relationships will be denoted as directed edges of the graph.

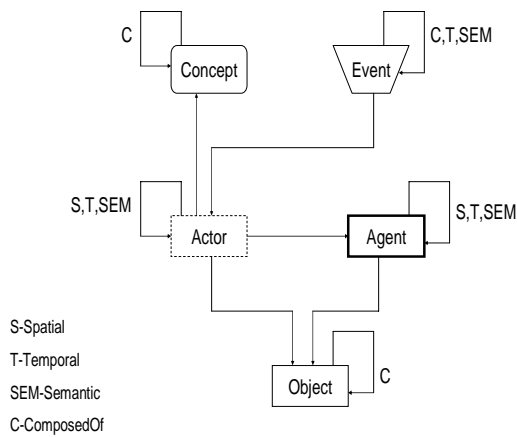

Figure1: Graphical representation of *DVSM*

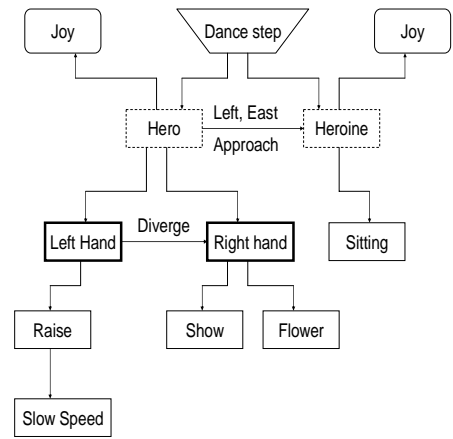

Figure2: Graph of dance step containing actors and agents.

The conceptual representation of an event highlighting a dance step as an instance of the graph, is depicted in Figure2. In this figure2, the dance event consists of two actors whose roles are hero and heroine. Hero is standing left to the heroine initially and facing east. Heroine is sitting and facing west. Now hero approaches the heroine. These spatio-temporal semantics are stored as relations. Event independent characteristics of the actor are stored as video objects separately (not shown in the figure). The actors express joy and it is initialized as emotion. Hero raises his left hand to chest level with medium speed and displays a flower to the heroine with his right hand. Heroine remains idle without performing any action. This dance step is choreographed as part of one line of lyrics of a song. Due to overflowing of nodes, attribute nodes are not shown in this figure. The entity classes and relationships of the model are formally defined as shown below:

### 4.1. Event Entity Class

Dance step of a song is known as a dance event. For instance, consider a Bharathanatyam step Samathristy (Saraswathi, 1994). It is performed with the eyes by keeping them static without blinking. This step represents a thought, firmness, surprise or an image of an angel. Also, dance events can be combined to form a composite dance event. As an illustration, consider a dance step, Chandran. This step represents a moon and is a combination of two other steps: Pathagam and Lola pathmam and should be rendered concurrently. Pathagam is performed by keeping the thumb closed and the other four fingers straight and denotes clouds, air, sword and blessing. Similarly, Lola pathmam is performed by keeping all the fingers open and stretched and represents a sun. Composite dance event many represent events which are rendered concurrently or sequentially by a dancer.

Dance events are composed of actors, posture of the actors, cognitive state of the actors and the interactions in space and time between agents and actors. Formally, a dance event is described as a tuple,

$$Event = \{ EID, D, AL, ND, ML \}$$

where, EID is a unique identifier of the dance step, D is the description of the dance step, AL denotes the list of actors, ND is the number of dancers who are performing steps in the event and ML is the media locator of the video clip. Here, this Event tuple corresponds to Event object of Figure1.

### 4.2. Basic Entity Class

A dance video object refers to a meaningful semantic entity of a dance video database. It can be described using attributes which can represent macro and micro features. Formally, it is defined as shown below:

$$Object = \{ OID, V, TY \}$$

where OID is a unique object id, V = {a1:v1, ..., an:vn } is *n* event independent or dependent attribute value pairs and TY = { AID, AGID } denotes the dependency of the object, either actor or agent (to be defined later).

For example, hero is showing a flower in his right hand. Here, *ShowFlower* is the object in which V denotes attributes action and instrument with values show and flower respectively. TY holds ID of the agent, Right Hand which belongs to the actor Hero. In this case, V represents event dependent (i.e., dance step dependent) values. Similarly, object may also represent any of the macro features.

### 4.3. Actor Entity Class

Actor is a spatiotemporal entity in dance videos. So the existence time (Vijay, 2004) can be associated with the entity and it represents the life span of it. Actor is also a spatial entity. Therefore actor's displacement in space is modeled using Trajectory Points as in MPEG-7 (Martinez, 2003). Hence, actors are spatio-temporal entities playing context dependent roles in the events. Actors can have spatial, temporal and event specific semantic attributes describing their roles. The roles can be linguistic roles (Martinez, 2003) as in MPEG-7 or any semantic roles, such as loves.

The existence time predicate ΦACTOR, which is associated with the actor entity class, defines life span of the actor in terms of the existence time granularity (e.g. min and sec). $\Phi ACTOR: S(ACTOR) \times Z \rightarrow B$. This predicate takes a particular actor entity and a particular granule (denoted by an integer; say sec) and evaluates to a Boolean. If it is true, then that actor exists in the modeled reality at that granule (sec).

Constraint.1: *Life span of an actor can exist only within the defined lifespan of the event to which it belong*.

Formally, an actor entity can be described as follows:

$$Actor = \{ AID, EID, DID, R, L, T, P \}$$

where EID is the event id, DID is the corresponding dancer id, R denotes either semantic or linguistic roles of an actor, L is the existence time or lifespan, T represents the trajectory points(Point Set) as in Mpeg7 and P is the posture of the actor, which is a basic entity.

### 4.4. Agent Entity Class

Agent entity class represents the finer spatio-temporal semantics of the actions. The agent entity is the one which is most important in dance videos. The essence of a dance step is the actions done by the actors and it is the agent that performs the action. This is an exclusive feature of the dance videos. All other video types possess just one or two agents, which are fixed and do not play any significant role at all. For example, legs are agents in soccer sport videos, bat and ball are agents in cricket sport videos. Agent entity elegantly models the action of the agent which belongs to an actor. For instance, left eye and right eye of a heroine are agents. Formally, it is defined as:

$$Agent = \{ AGID, AID, EID, L, T, X, S, I \}$$

where AID and EID denote the actor id and event id respectively, X is the action agent performs, S denotes speed of X and I is the instrument held by the agent. Also, L and T depict the lifespan and spatial trajectory, similar to actor objects. Here, X, S and I are all basic entity types as defined earlier.

### 4.5. Concept Entity Class

The cognitive and affective content of an actor is modeled as a concept object. The concept is modeled as a separate entity type because of its ontological nature, thereby improving the semantic search. Formally, a concept entity can be defined as

$$Concept = \{ CID, AID, EID, T, D \}$$

where T = { Emotion, Feeling, Mood } and D denote type of the concept and description as a string using natural language respectively.

### 4.6. Interaction Relationships

An interaction relationship relates members of an entity set to those of one or more entity sets. The DVSM employs the following set of relationships between the different entity sets. They are Composition (C), Spatial(S): which are topological (Egenhofer, 1994) and directional (Li, 1996), Temporal(T): Allen's interval algebra (Allan, 1983), Spatio-temporal, Motion(M): such as approach, diverge, stationary which are defined over the basic temporal relations (Athanasios, 2005), Semantic(SE) and Ontological(O).

The following section describes the set of relationships that occur between the various dance video entity sets.

### 4.6.1. Event Relationship

The relations between events are composition and temporal. Intuitively, a dance step of an actor may be followed by another actor immediately. Similarly, a dance step of an actor may be repeated by another dancer some time later. These follows and repeats relations are cues for later retrieval and mining operations. For example the query, find the set of dance steps done by a dancer, that is repeated by another dancer, can be processed by checking the life spans of the corresponding events.

Suppose $E1, E2, ..., En$ are dance events participating in a temporal relationship. Let $a1$ and $a2$ be the actors, $x1$ and $x2$ be the actions of agents present in $E1$ and $E2$ respectively. Then, the predicate $\Phi REPEATS: S(X) \times S(A) \rightarrow E$ can take an actor and action and can return a set of events in which the action is performed. There is a constraint on the REPEATS predicate.

**Constraint.2**. *Let LS1 and LS2 be the lifespan of E1 and E2 respectively. Then*

$$(x1 = x2) \vee (LS1 < LS1) \Longrightarrow (E1 = E2)$$

Similarly, the other predicates such as *performSameStep, performDifferentStep*, and *observe* can be formulated, apart from *follows* and *repeats* predicates. Event relationships are formally defined as follows:

$$EE = \{ SRC, TAR, LST \}$$

where SRC and TAR denote the source and target event ids and LST is the set of composition and temporal relationships which hold between source and target events.

### 4.6.2. Object Relationship

Objects can be composed of other objects. For example, consider Figure2 where hero holds a flower in his right hand. Here, flower is an example of an object. Formally, the relationship between objects can be represented similar to event relationships with a restriction that the SRC and TAR can be basic entities and LST will contain only composition relations.

### 4.6.3. Actor Relationship

Actor relationship represents the relationship between the roles of the objects, such as relation between hero and heroine who are dancer objects. Spatial, temporal and semantic relationships exist between the actors in a particular dance event. For instance, hero standing left to the heroine initially, may approach the heroine. This dance semantic contains spatial and motion relationships left and approach respectively. The actor relationship is formally defined as shown below:

$$AA = \{ AID1, AID2, O1, O2, LST \}$$

where AID1 and AID2 are roles of the dancers O1 and O2 respectively and LST is now the set containing spatial, temporal and semantic relationships. Note that O1 and O2 are basic entity types.

### 4.6.4. Agent Relationship

Agent relationship is a second level semantic relation that describes the spatial and temporal characteristics of the agents. That is, agent relationship represents the finer semantics between the body parts of an actor. For instance, heroine is touching her left cheek with the index finger of her right hand. So, left cheek and right hand are the agents and finger can be the instrument used in the semantic relationship touch. Agent relationship is formally defined as,

$$AGAG = \{ AGID1, AGID2, AID, LST \}$$

where AGID1 and AGID2 are agentIDs of an actor AID and LST is similar to actor relationships.

### 4.6.5. Concept Relationship

Concept relationship is an ontological relationship (O) between concept entities. Typical ontological relationships (Guiness, 2004) are subClassOf, cardinality, intersection and union. This relationship is similar to event relationship with a modification that the source and target ids represent concepts and LST holds only ontological relations. All other types of relationships between the different dance video entities are either semantic relationships or composition relationships such as partOf, composedOf, memberOf and so on. Table 1 summarizes the semantics of the kinds of relationships that exist between the dance video entities.

Table 1. Semantics of Relationships.

|         | Event  | Object | Actor  | Agent  | Concept |
|---------|--------|--------|--------|--------|---------|
| Event   | C,T,SE |        | C      |        | C       |
| Object  |        | C      | C      | C      |         |
| Actor   | C      | C      | S,T,SE | C      | C       |
| Agent   |        | C      | C      | S,T,SE |         |
| Concept | C      |        | C      |        | O       |

## 5. Implementation of DVSM

We have implemented the model in order to annotate the macro and micro features that are associated with the dance video. The tool has been developed using J2SE1.5 and JMF2.1.1 under Dell workstation. The tool is interactive as it minimizes the hard coding.

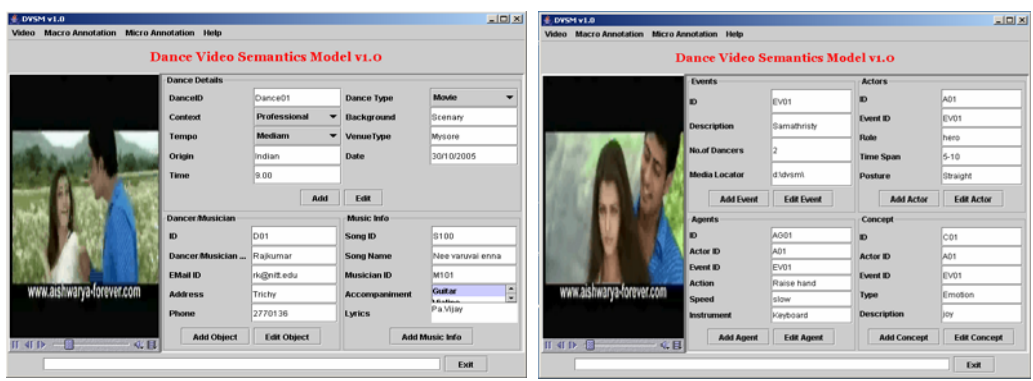

Fig.3 and Fig.4. Macro annotation and Micro annotation of the event semantics.

The dance video can be annotated by looking at the video clips that is running. Macro features can be annotated initially. The details of the dancers, musician, music, song, background, tempo, dance origin, context (whether live, rehearsal, professional play, competition etc), date and time of recording, type of performance venue and type of dance video are annotated. The screen shot depicting the rendering of the dance and interactive annotation of macro features is shown in Figure 3.

Then, micro features of every dance step of a song have to be annotated. The screen shot depicted in Figure 4 represents events, actors, agents and concepts. The annotator, by looking at the video, annotates the different information pertaining to these entity types in the order: event, actors of this event, agents of the actors, concepts revealed by the actors. But, one can swap the annotation of agents and concepts depending on his interest. The user interface has been carefully designed such a way that it minimizes the hard coding, as many of the graphical components will be populated automatically.

The second part of the micro features annotation involves the description of the various relationships between the entity types. For instance, event relationships, actor relationships, agent relationships and concept relationships describe the spatial, temporal, motion and semantic relations that exist between the entity types. The annotated data are stored in a backend database.

## 6. Evaluation

Batini(1996) posits that conceptual model should possess the basic qualities: expressiveness, simplicity, minimality and formality. Additionally, Harry(2001) outlines other semantic qualities for video types. They are: explicit media structure(M), ability to describe objects (O), events (E), spatial relationships(S), temporal relationships(T) and integration of syntactic and semantic information (I). This paper introduces another factor, contextual description (Actor(A) and Agent(G)) to evaluate the proposed model.

The DVSM satisfies all the semantic quality requirements. Moreover, DVSM is unique in modeling the finer spatio-temporal contextual semantics of the events at finer granularity, with the help of agent entity type. Table 2 contrasts the existing semantic content based models against the semantic quality factors. The table illustrates that some models lack semantic features, some lack syntactic features and only few models integrate both syntactic and semantic relationships. Some applications, like soccer sports video, require the model to represent the contextual features of objects (called actors). The ESSVM proposed by Ahmet et al, describes contextual information at actor level. However, dance videos require contextual description at multiple granularities (called agent), beyond the actor level. Our proposed semantic model possesses both contextual abstractions-actor and agent, apart from the other semantic qualities. Hence, with the agent based approach, the paper claims to have achieved conceptual, semantic and contextual qualities in dance video data modeling.

Table 2. Comparison of semantic video models.

Legend: M-Media structure, O-Object, E-Event, S-Spatial, T-Temporal, I-Syntactic-semantic info, A-Actor, G-Agent.

| Model | Semantic Qualities | | | | | | | |
|---|---|---|---|---|---|---|---|---|
| | M | O | E | S | T | I | A | G |
| AVIS | × | × | | × | × | × | | |
| OVID | × | × | | × | × | × | | |
| QBIC | × | × | | × | | | | |
| DISIMA | × | × | × | × | × | | | |
| COSMOS7 | | × | × | × | × | × | | |
| ATN | × | × | | × | × | × | | |
| ESSVM | × | × | × | × | × | × | × | |
| DVSM | × | × | × | × | × | × | × | × |

## 7. Conclusion

Data semantics provides a connection from a database to the real world outside the database and the conceptual model provides a mechanism to capture the data semantics (Vijay, 2004). The task of conceptual modeling is crucial and important, because of the vast amount of semantics that exist in multimedia applications. In particular, dance videos possess several interesting semantics for modeling and mining. This paper described as agent based approach for elicitation of the semantics such as macro and micro features of the dance videos. An interactive annotation tool has been developed based on the DVSM for annotating the dance video semantics at syntactic, semantic and contextual levels. Since dance steps are annotated manually, it is somewhat tedious to annotate dances by the dance expert.

Further work would be useful in many areas. It would be interesting to explore how DVSM can be used as a video model for exact and approximate query processing. As MPEG-7 is used to document the video semantics recently, it is valuable to employ MPEG-7 for representing dance semantics to enable better interoperability. Finally, it will be useful to explore how video mining techniques can be applied to dance videos.

## Acknowledgements

This work is supported fully by the University Grants Commission (UGC), Government of India grant *XTFTNBD065*. The authors thank the editors and the anonymous reviewers for their insightful comments.

### Appendix-A: List of Agents

| | | | | | |
|---|---|---|---|---|---|
| Head | Hand | Knee | Leg | Foot | Arm |
| Finger | Ankle | Elbow | Heel | Lower Leg | Wrist |
| Toe | Hip | Shoulder | Waist | Back | Torso |
| Forearm | Palm | Pelvis | Thigh | Ball of Foot | Chest |

### Authors' Information

**Balakrishnan Ramadoss** - Assistant Professor in the Department of Computer Applications, National Institute of Technology, Tiruchirappalli, India (Deemed University).

email: brama@nitt.edu.

**Kannan Rajkumar** - M.Phil, Lecturer of Computer Science at Bishop Heber College, Tiruchirappalli. Currently he is pursuing Ph.D. in Computer Applications at National Institute of Technology, Tiruchirappalli. e-mail: rajkumarkannan@yahoo.co.in